\documentclass[conference]{IEEEtran}
\IEEEoverridecommandlockouts
\usepackage{cite}
\usepackage{amsmath,amssymb,amsfonts}
\usepackage{algorithmic}
\usepackage{graphicx}
\usepackage{textcomp}
\usepackage{xcolor}
\usepackage{booktabs}
\usepackage{multirow}
\usepackage{graphicx}      
\usepackage[table]{xcolor} 
\usepackage{placeins}
\usepackage{url}
\usepackage[hidelinks]{hyperref}

\def\BibTeX{{\rm B\kern-.05em{\sc i\kern-.025em b}\kern-.08em
    T\kern-.1667em\lower.7ex\hbox{E}\kern-.125emX}}
\begin{document}

\title{Drum Synthesis from Expressive Drum Grids via Neural Audio Codecs%
\thanks{Project page: \url{https://github.com/kostantinos-soiledis/midigroove_poc}}}


\author{
\IEEEauthorblockN{
Konstantinos Soiledis\IEEEauthorrefmark{1}\IEEEauthorrefmark{2},
Maximos Kaliakatsos-Papakostas\IEEEauthorrefmark{1},
Dimos Makris\IEEEauthorrefmark{1},
Konstantinos Tsamis\IEEEauthorrefmark{1}\IEEEauthorrefmark{2}
}

\IEEEauthorblockA{\IEEEauthorrefmark{1}
Dept. of Music Technology and Acoustics, Hellenic Mediterranean University,
Rethymno \& Athens, Greece}

\IEEEauthorblockA{\IEEEauthorrefmark{2}
Athena RC, Athens, Greece}

\IEEEauthorblockA{
k.t.soiledis@gmail.com}
}
\maketitle

\begin{abstract}
Generating realistic drum audio directly from symbolic representations is a challenging task at the intersection of music perception and machine learning. We propose a system that transforms an expressive drum grid (a time-aligned MIDI representation with microtiming and velocity information) into drum audio by predicting discrete codes of a neural audio codec. Our approach uses a Transformer-based model to map the drum grid input to a sequence of codec tokens, which are then converted to waveform audio via a pre-trained codec decoder. We experiment with multiple state-of-the-art neural codecs i.e. EnCodec, DAC, and X-Codec to assess how the choice of audio representation impacts the quality of the generated drums. The system is trained and evaluated on the Expanded Groove MIDI Dataset (E-GMD), a large collection of human drum performances with paired MIDI and audio. We evaluate the fidelity and musical alignment of the generated audio using objective metrics. Overall, our results establish codec-token prediction as an effective route for drum grid–to–audio generation and provide practical insights into selecting audio tokenizers for percussive synthesis
\end{abstract}


\section{Introduction}
Automating the transformation of symbolic drum representations into realistic audio could benefit music production and deepen our understanding of cross-modal learning. Conventional MIDI-to-audio rendering typically relies on sample libraries or drum synthesizers, which can reproduce patterns effectively but often fail to capture the nuance of human performance, such as microtiming deviations and dynamic variation (i.e. velocity) that shape perceived ``feel.'' Motivated by recent progress in neural audio generation, we study \textbf{drum performance rendering} as a conditional generation problem: given an \textbf{expressive drum grid} (a time-aligned MIDI-derived representation encoding timing deviations and hit strength), generate the corresponding \textbf{audio waveform} of the performed groove. We train and evaluate on the Expanded Groove MIDI Dataset (E-GMD)~\cite{b3}, which contains 444 hours of aligned drum audio and MIDI across 43 kits and includes detailed velocity annotations, enabling large-scale supervised learning for this inverse task (audio generation from MIDI).

Our approach follows the codec-token paradigm used in systems such as AudioLM and MusicLM~\cite{b1,b2}. Rather than predicting waveform samples directly, we predict discrete token sequences from a pretrained neural audio codec and decode them back to waveform. This delegates fine-scale waveform reconstruction to the fixed codec decoder while the model learns a structured mapping from grid features to audio tokens. A key question we investigate is whether \emph{codec choice} affects conditional generation quality. We compare three state-of-the-art neural codecs—EnCodec~\cite{b4}, DAC~\cite{b6}, and X-Codec~\cite{b5}—which differ in bitrate, reconstruction fidelity, and (for X-Codec) the use of semantic-side objectives intended to mitigate shortcomings of purely acoustic tokenizations. To our knowledge, this provides one of the first controlled comparisons of these codec token spaces for symbolic-conditioned drum audio generation.

\textbf{Contributions:} This paper makes the following contributions:
\begin{itemize}
    \item \textbf{Grid-to-audio generation model:} We introduce a Transformer-based system that maps a fully expressive drum grid to drum audio by predicting neural codec tokens and decoding them to waveform.
    \item \textbf{Neural codec comparison:} We perform an empirical comparison of EnCodec, DAC, and X-Codec as intermediate representations for drum audio generation, isolating the impact of codec choice under a controlled modeling setup.
\end{itemize}

We evaluate on objective metrics assessing fidelity and timing alignment, providing a multifaceted view of reconstruction. Additional qualitative examples (generated audio) and extended results/plots, including the full all-kits evaluation breakdown, are provided on the project page.

\section{Related Work}

\textbf{Drum Audio Synthesis and Generation:}
Early work on learned drum audio generation often treats drums as a \emph{one-shot} synthesis problem, generating isolated kick/snare/cymbal samples rather than time-synchronized performances. DrumGAN is a representative example, using a GAN to synthesize single-hit drum sounds with perceptually meaningful timbral conditioning; the model is effective for sound design and timbre exploration, but it is not conditioned on a bar-level rhythmic context and therefore does not directly address rendering an entire groove as audio \cite{b7}.
On the symbolic side, expressive groove modeling has been studied with models such as GrooVAE, which learn microtiming and velocity patterns from human performances in the Groove MIDI Dataset (GMD) \cite{b8}. These systems generate realistic MIDI-level performances (“when” and “how hard”), but they do not produce audio waveforms.

\textbf{MIDI/Drum-Grid to Audio via Learned Encoders:}
Bridging symbolic control to waveform audio has traditionally relied on sample libraries or conventional drum synthesizers. A more modern direction learns the rendering function directly from paired symbolic--audio data. A practical two-stage neural synthesizer for \emph{MIDI-to-audio} synthesis is proposed in \cite{b9}. A Transformer encoder-decoder maps MIDI to spectrograms, and a neural spectrogram inverter converts spectrograms to waveform audio . This work is important because it demonstrates an explicit \emph{symbolic encoder} paired with a learned acoustic decoder target, offering note-level control and real-time generation.
For drum-specific audio generation, CRASH conditions a score-based diffusion model on structured score information to generate high-resolution raw audio drum sounds and enables controllable sampling operations (e.g., inpainting and interpolation) \cite{b10}. CRASH targets short percussive sounds rather than multi-bar performance rendering, but it establishes that structured “score/grid” control is a viable conditioning interface for high-fidelity drum audio.

\textbf{Neural Audio Codecs and Token-Based Generative Models:}
Neural audio codecs enable high-fidelity reconstruction while mapping audio into compact discrete sequences. SoundStream introduced end-to-end neural coding with residual vector quantization (RVQ), and EnCodec extended this line with high-fidelity neural audio compression that has become a standard tokenizer for audio generation \cite{b11,b4}. DAC improves RVQ-based codec learning and reconstruction quality, making it an appealing target representation for generative modeling \cite{b6}. Building on these discrete representations, AudioLM frames audio generation as language modeling over codec tokens, showing that token modeling can preserve local fidelity while supporting longer-range structure \cite{b1}. MusicLM extends the token paradigm to text-conditioned music generation \cite{b2}. In our setting, codec tokens are used not for unconditional continuation, but as a supervised target for \emph{symbolic-conditioned} rendering from expressive drum grids.

\textbf{Codec-LM MIDI-to-Audio and Drum Generation with Rhythm Control:}
Recent work has begun to apply codec-token modeling directly to symbolic-to-audio synthesis. MIDI-VALLE adapts VALL-E-style neural codec language modeling to performance MIDI-to-audio synthesis (shown for expressive piano), conditioning on symbolic performance inputs and a reference prompt to control acoustics/timbre \cite{b12}. Although not drum-focused, it provides a close precedent for treating symbolic-to-audio synthesis as conditional codec-token prediction.
In parallel, several drum-generation systems emphasize \emph{rhythm control} but use audio-form prompts rather than MIDI. STAGE fine-tunes MusicGen (which uses EnCodec tokens) for stemmed accompaniment generation \cite{b13}, and DARC extends this approach to drums by adding fine-grained rhythm control from explicit rhythm prompts such as tapping or beatboxing while conditioning on musical context \cite{b14}. TRIA proposes masked token modeling over DAC tokens to map arbitrary percussive gestures into high-fidelity drum recordings, conditioning separately on rhythm and timbre prompts \cite{b15}. These methods demonstrate strong drum rendering quality in codec-token spaces, but their conditioning interfaces are primarily audio prompts and/or stem context, whereas our work focuses on rendering from an explicit \emph{expressive drum grid} derived from MIDI.

\section{Methodology}

\subsection{Overview}
Our goal is to render realistic drum audio from a symbolic performance. The system takes as input an expressive drum grid derived from MIDI and outputs an audio waveform by predicting discrete neural audio codec tokens. The approach has three stages: (i) constructing paired supervision by aligning MIDI-derived grids with codec tokens extracted from the paired audio, (ii) training a Transformer encoder to map grids (plus metadata) to token sequences, and (iii) synthesizing waveforms by decoding the predicted tokens with the corresponding pretrained codec decoder. All experiments use the Expanded Groove MIDI Dataset (E-GMD) as the source of aligned MIDI and audio \cite{b3}, and we treat the codec models (EnCodec, DAC, X-Codec) as fixed tokenizers/decoders \cite{b4,b6,b5}.

\subsection{Paired supervision via cached 4-beat windows}
We preprocess E-GMD by cutting each performance into beat-aligned, 4-beat segments. For each segment, we take (i) the exact audio slice for those four beats and (ii) the MIDI events that occur in the same time span, and we align both onto the same frame grid. We then cache two paired, frame-synchronous objects per segment:
(1) an “expressive grid” computed from the MIDI, and
(2) the target sequence of codec tokens produced by running the segment’s audio through a fixed, pretrained codec encoder.

Because segments are defined in beats, their length in seconds depends on the local tempo. Alongside the tensors, each cached item stores the information needed to re-extract the identical slice (start time, duration, performance/track IDs or paths, and native sample rate). Using this fixed cache means every experiment trains and evaluates on the same audio/MIDI slices and the same codec targets, making comparisons across systems consistent.

\subsection{Expressive drum grid representation}
\begin{figure}[t]
    \centering
    \includegraphics[width=\columnwidth]{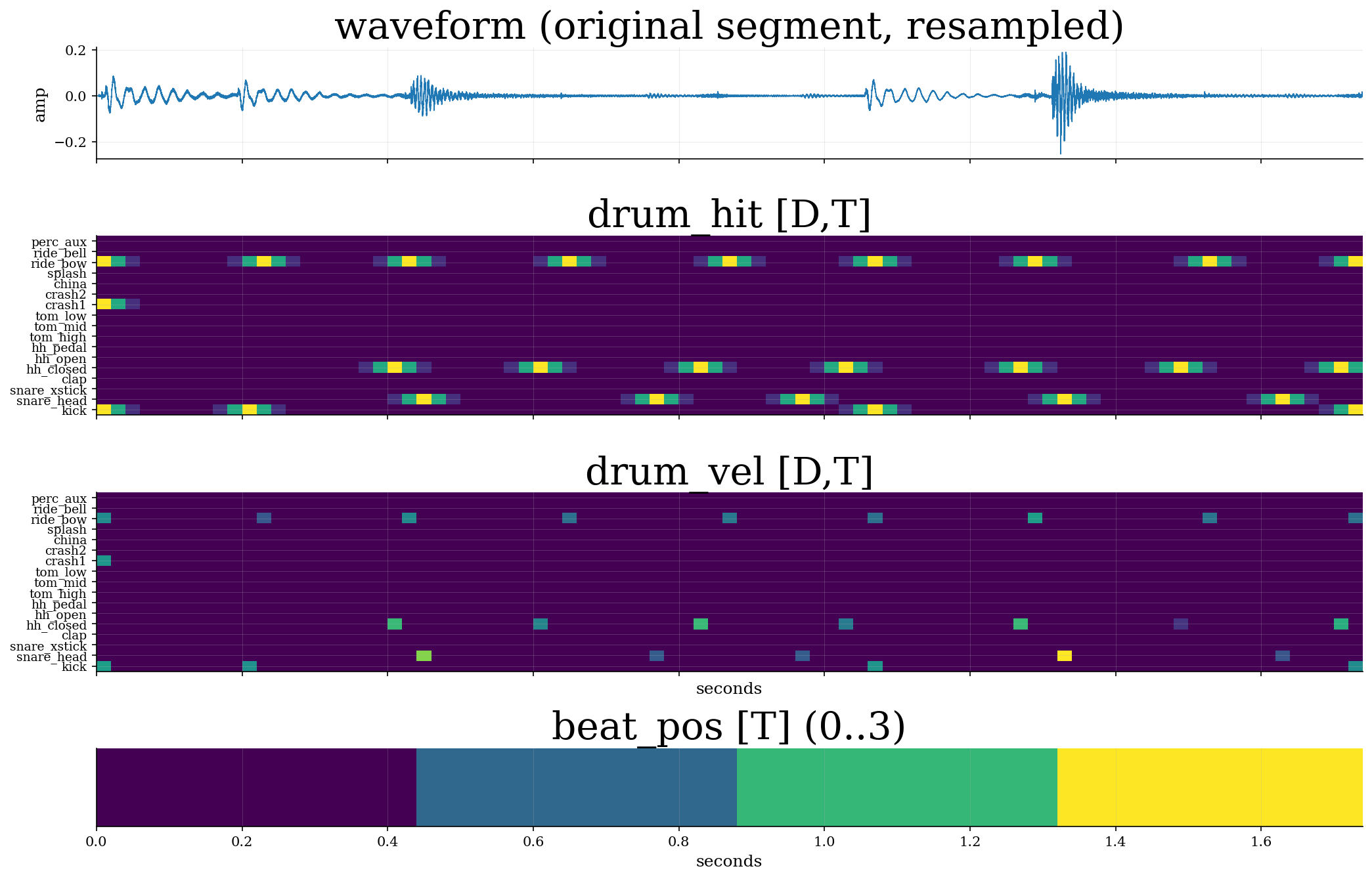}
    \caption{Example expressive drum grid representation (hit strength, velocity-at-onset, and beat position).}
    \label{fig:grid}
\end{figure}

For each training window, we convert the MIDI performance into a multi-lane grid on the same temporal frame grid as the codec token sequence for that window. Let $T$ be the number of codec token frames in the window and let $\mathrm{fps}=T/\texttt{window\_seconds}$ be the corresponding effective frame rate (about 50~fps for EnCodec/X-Codec in our setup and about 86--88~fps for DAC). We index drum lanes by $d \in \{1,\ldots,D\}$ (kick, snare, toms, hi-hat variants, cymbals, etc.), using separate lanes for distinct articulations when available (Fig.~\ref{fig:grid}).

We construct two core per-frame lanes:
\begin{itemize}
  \item \textbf{Hit strength} $\texttt{drum\_hit}[d,t]$: a real-valued hit-activity signal. Each onset at time $\tau$ (relative to the window start) is mapped to the nearest frame index $t_0=\mathrm{round}(\tau\,\mathrm{fps})$, and a fixed-width Gaussian bump centered at $t_0$ is written over neighboring frames, taking the elementwise maximum when bumps overlap. This yields a smooth, frame-aligned onset representation.
  \item \textbf{Velocity-at-onset} $\texttt{drum\_vel}[d,t]$: normalized MIDI velocity placed at the onset frame $t_0$; non-onset frames are zero.
\end{itemize}

All enabled lanes are concatenated into a feature tensor $\texttt{grid}[F,T]$, where $F$ is the total number of lanes. In addition to per-frame features, the model conditions on window-level metadata (tempo $\texttt{bpm}$, performer identity $\texttt{drummer\_id}$, and optionally $\texttt{kit\_name\_id}$). We also provide a beat-phase sequence $\texttt{beat\_pos}[t]\in\{0,1,2,3\}$ giving the beat index modulo~4, computed from $\mathrm{fps}$ and $\texttt{bpm}$ relative to the window start.

\subsection{Codec token targets}
For each window, the paired audio segment is encoded with a pretrained neural audio codec to obtain a discrete supervision target
$\texttt{tgt}[c,t]$, where $c \in \{1,\dots,C\}$ indexes codebooks and $t \in \{1,\dots,T\}$ indexes codec frames. Although codec families differ in their quantizer structure (number of codebooks and codebook cardinality), the supervised objective is the same in all cases: predict the aligned token tensor $\texttt{tgt}[C,T]$ from the expressive drum grid.

Across all experiments, codec parameters are held fixed and we vary only the grid-to-token predictor. For \textbf{EnCodec}, we use the pretrained \texttt{facebook/encodec\_32khz} checkpoint, which is commonly used as a music tokenizer with four RVQ codebooks of size 2048 per frame.
For \textbf{DAC}, we use the pretrained \texttt{descript/dac\_44khz} checkpoint, which employs $C{=}9$ residual codebooks with codebook size 1024; we treat DAC as a high-fidelity, domain-general RVQ codec baseline for waveform tokenization and reconstruction.
For \textbf{X-Codec}, we use \texttt{hf-audio/xcodec-hubert-general} and select a bandwidth of 2.0\,kbps, which corresponds to an effective token tensor with $C{=}4$ codebooks per frame in this configuration (while higher bandwidths increase $C$).
We include X-Codec because it explicitly incorporates semantic features from a self-supervised encoder (e.g., HuBERT) into the codec and introduces an auxiliary semantic reconstruction objective, potentially yielding token targets with different structural and perceptual properties than purely acoustic codecs.

\subsection{Transformer token prediction model}
Given a cached 4-beat window, our model predicts the codec token sequence directly from the MIDI-derived grid, one frame at a time, using a non-autoregressive Transformer encoder.

For each frame $t$, the per-frame grid vector $\texttt{grid}[:,t]$ is first projected to the Transformer dimension with a learned linear layer. We then \emph{add} (i) a learned embedding of the beat position within the 4-beat window, $\texttt{beat\_pos}[t]$, and (ii) a learned absolute position embedding for frame $t$. Global conditioning is incorporated as an additive per-window bias that is broadcast to every frame: we project a log-scaled tempo feature $\log(1+\texttt{bpm})$ and add learned embeddings for $\texttt{drummer\_id}$ and, when available, $\texttt{kit\_name\_id}$. The resulting sequence of per-frame vectors is passed through bidirectional self-attention over all $T$ frames in the window, so each frame can use context from the entire 4-beat segment.

The Transformer outputs contextual states $\mathbf{h}_t$ for all frames. A linear head maps each $\mathbf{h}_t$ to logits over the codec vocabulary for each codebook at that frame, and we train with mean cross-entropy over all codebooks and frames (ignoring PAD). This setup learns to ``render'' the window by using full-window context to capture rhythmic structure and timbral changes without autoregressive decoding.

\subsection{Training and inference}
We train a separate token predictor for each codec family (EnCodec, X-Codec, DAC) using the paired windows described above. All reported runs use beat-synchronous 4-beat windows and restrict to $\texttt{beat\_type}=\texttt{beat}$. We evaluate two dataset settings: (i) \textbf{single-kit}, where training windows are filtered to $\texttt{kit\_category}=\texttt{Acoustic\ Kit}$, and (ii) \textbf{all-kits}, where we include all kits except a small excluded set (Custom1, Custom2, Custom3). All models were trained on a single NVIDIA GeForce RTX 3080 GPU (10\,GB VRAM).

\paragraph{Model variants.}
All systems share the same non-autoregressive Transformer encoder formulation, differing only in capacity:
\begin{itemize}
  \item \textbf{Base:} $d_{\text{model}}=768$, $L=6$ layers, $H=8$ attention heads, feedforward multiplier $=4$, dropout $=0.1$.
  \item \textbf{Large:} $d_{\text{model}}=1536$, $L=10$ layers, $H=12$ attention heads, feedforward multiplier $=4$, dropout $=0.1$.
\end{itemize}
Inputs use $\texttt{drum\_hit}$ and $\texttt{drum\_vel}$ lanes. Global conditioning includes log-scaled BPM (using $\log(1+\texttt{bpm})$) and a learned drummer embedding. For \textbf{all-kits} experiments we additionally condition on kit identity via a learned $\texttt{kit\_name\_id}$ embedding; in the \textbf{single-kit} setting kit identity is constant and omitted.

\paragraph{Batching and masking.}
Beat-synchronous windows have variable duration in seconds, and thus variable token length $T$. Within each minibatch, sequences are padded to the maximum $T$ and targets are padded using the codec-specific PAD id. The loss ignores padded positions via $\texttt{ignore\_index}=\texttt{pad\_id}$, and the Transformer is provided a corresponding validity mask so that self-attention does not attend to padding tokens.

\paragraph{Optimization and model selection.}
Training uses AdamW with a constant learning rate of $6\times 10^{-5}$, and global gradient clipping at 1.0. We set up a max of 200{,}000 optimizer steps and run validation every 300 steps. Early stopping is governed by validation NLL, terminating after 5000 steps without improvement. \textbf{Base} models use batch size 24 and \textbf{Large} models use batch size 8.

\subsection{Evaluation protocol and metrics}
Evaluation is designed to compare multiple systems on a standardized set of windows. When comparing across codecs, we restrict scoring to the \emph{intersection} of windows present in all systems, identified by stable metadata-derived keys, ensuring that all models are evaluated on identical segments. For the single-kit configuration, this yields 1748 evaluation windows; for the all-kits configuration we evaluate on up to 68180 windows.

We report both token-level and audio-level metrics:
\begin{itemize}
    \item \textbf{Token metrics:} negative log-likelihood (NLL), perplexity, and overall token accuracy (all computed with PAD ignored). For perplexity, we compute per-window perplexity as $\exp(\mathrm{NLL}_w)$ where $\mathrm{NLL}_w$ is the mean token NLL within window $w$, and report mean$\pm$std over windows (so this does not generally equal $\exp(\mathrm{mean\ NLL})$).\footnote{In separate diagnostic summaries that report entropy in bits/token, perplexity is defined as $2^H$; we distinguish this from the $\exp(\mathrm{NLL})$ definition used for the main evaluation tables.}

    \item \textbf{Audio metrics:} predicted tokens are decoded, resampled to a common evaluation sample rate (32~kHz), and compared to reference audio using sample-level RMSE and MAE, multi-resolution STFT spectral convergence ($\texttt{mr\_stft\_sc}$), transient-to-tail energy ratio error ($\texttt{tter\_db\_mae}$), and RMS-envelope correlation ($\texttt{env\_rms\_corr}$). MR-STFT spectral convergence averages spectral-convergence terms over three STFT resolutions with $(N_{\mathrm{FFT}}, \mathrm{hop}) \in \{(512,128), (1024,256), (2048,512)\}$; all audio metrics are computed per window and then averaged.

    \item \textbf{Onset alignment:} we compute onset precision, recall, and F1 by matching onsets detected in the decoded predicted audio to reference onsets derived from the cached expressive drum grid, using a $\pm 50\,\mathrm{ms}$ tolerance window. Reference onsets are obtained by taking frames where the maximum onset-velocity across drum lanes exceeds a threshold $\left(\max_d \texttt{drum\_vel}[d,t] > 0.30\right)$ and converting those frame indices to times using the cached window frame rate. Predicted onsets are extracted from the synthesized waveform using the \texttt{madmom} onset detection pipeline. Matches are counted within the tolerance to measure whether generated transients align with the intended rhythmic events.

    \item \textbf{Perceptual distance:} Fr\'echet Audio Distance computed in a learned embedding space using the CLAP-LAION-music audio encoder (via \texttt{fadtk})~\cite{b16,b17}. We embed decoded predictions and aligned reference windows (same segmentation and durations as above) and compute the Fr\'echet distance between the embedding distributions (lower is better). To reduce sample-size bias, we report the extrapolated variant $\mathrm{FAD}_\infty$ for the single-kit.
\end{itemize}

\begin{table*}[!t]
\centering
\caption{Evaluation on one-kit and all-kits test sets (Base models).}
\renewcommand{\arraystretch}{1}
\setlength{\tabcolsep}{1pt}
\resizebox{\textwidth}{!}{%
\begin{tabular}{|p{0.060\linewidth}|p{0.060\linewidth}|p{0.060\linewidth}|p{0.060\linewidth}|p{0.060\linewidth}|p{0.07\linewidth}|p{0.07\linewidth}|p{0.11\linewidth}|p{0.11\linewidth}|p{0.12\linewidth}|p{0.067\linewidth}|p{0.067\linewidth}|p{0.067\linewidth}|p{0.064\linewidth}|}
\hline
\textbf{Eval} & \textbf{Codec} & \multicolumn{3}{c|}{\textbf{Token metrics}} & \multicolumn{5}{c|}{\textbf{Audio metrics}} & \multicolumn{3}{c|}{\textbf{Onset metrics}} & \\
\cline{3-5}\cline{6-10}\cline{11-13}
\textbf{Setting} & \textbf{} & \shortstack[c]{\scriptsize \textbf{\textit{NLL}$^{\mathrm{a}}$\,\,$\downarrow$}} & \shortstack[c]{\scriptsize \textbf{\textit{PPL}$^{\mathrm{a}}$\,\,$\downarrow$}} & \shortstack[c]{\scriptsize \textbf{\textit{Acc(\%)}$^{\mathrm{a}}$\,\,$\uparrow$}} & \shortstack[c]{\scriptsize \textbf{\textit{RMSE}$^{\mathrm{b}}$\,\,$\downarrow$}} & \shortstack[c]{\scriptsize \textbf{\textit{MAE}$^{\mathrm{b}}$\,\,$\downarrow$}} & \shortstack[c]{\scriptsize \textbf{\textit{MR-STFT SC}$^{\mathrm{b}}$\,\,$\downarrow$}} & \shortstack[c]{\scriptsize \textbf{\textit{Env RMS corr}$^{\mathrm{b}}$\,\,$\uparrow$}} & \shortstack[c]{\scriptsize \textbf{\textit{TTER (dB) MAE}$^{\mathrm{b}}$\,\,$\downarrow$}} & \shortstack[c]{\scriptsize \textbf{\textit{P(\%)}$^{\mathrm{c}}$\,\,$\uparrow$}} & \shortstack[c]{\scriptsize \textbf{\textit{R(\%)}$^{\mathrm{c}}$\,\,$\uparrow$}} & \shortstack[c]{\scriptsize \textbf{\textit{F1(\%)}$^{\mathrm{c}}$\,\,$\uparrow$}} & \shortstack[c]{\scriptsize \textbf{\textit{FAD}$^{\mathrm{d}}$\,\,$\downarrow$}} \\
\hline
 OneKit & encodec & \cellcolor{green!15} $2.142\pm0.681$ & \cellcolor{green!15} $10.8\pm8.3$ & \cellcolor{green!15} $42.7\pm13.7$ & $0.0201\pm0.0106$ & $0.0100\pm0.0058$ & \cellcolor{green!15} $0.842\pm0.160$ & \cellcolor{green!15} $0.690\pm0.228$ & \cellcolor{green!15} $1.29\pm1.22$ & \cellcolor{green!15} $78.3\pm15.5$ & \cellcolor{green!15} $68.4\pm17.8$ & \cellcolor{green!15} $71.0\pm13.3$ & \cellcolor{green!15} $0.281$ \\
\hline
 OneKit & xcodec & $4.422\pm0.590$ & $102.1\pm81.8$ & $11.9\pm3.9$ & $0.0305\pm0.0176$ & $0.0161\pm0.0110$ & $1.357\pm0.702$ & $0.552\pm0.242$ & $1.92\pm1.60$ & $76.6\pm17.6$ & $64.8\pm16.5$ & $67.8\pm12.4$ & $0.350$ \\
\hline
 OneKit & dac & $6.265\pm0.450$ & $563.5\pm236.7$ & $3.8\pm6.0$ & \cellcolor{green!15} $0.0184\pm0.0094$ & \cellcolor{green!15} $0.0095\pm0.0054$ & $0.982\pm0.085$ & $0.580\pm0.235$ & $1.44\pm1.24$ & $69.8\pm15.1$ & $65.0\pm18.1$ & $65.0\pm12.3$ & $0.545$ \\
\hline
\hline
 AllKits & encodec & \cellcolor{green!15} $2.153\pm0.743$ & \cellcolor{green!15} $11.6\pm11.0$ & \cellcolor{green!15} $43.4\pm14.7$ & $0.0200\pm0.0118$ & $0.0103\pm0.0070$ & \cellcolor{green!15} $0.827\pm0.171$ & \cellcolor{green!15} $0.710\pm0.220$ & \cellcolor{green!15} $1.47\pm1.30$ & $78.2\pm15.8$ & \cellcolor{green!15} $68.2\pm18.2$ & \cellcolor{green!15} $70.6\pm13.4$ & \cellcolor{green!15} $0.193$ \\
\hline
 AllKits & xcodec & $4.429\pm0.617$ & $104.9\pm90.0$ & $12.5\pm4.3$ & $0.0336\pm0.0203$ & $0.0176\pm0.0125$ & $1.669\pm0.925$ & $0.568\pm0.252$ & $2.22\pm1.89$ & \cellcolor{green!15} $78.9\pm16.5$ & $64.4\pm16.7$ & $68.8\pm12.7$ & $0.277$ \\
\hline
 AllKits & dac & $6.153\pm0.540$ & $521.8\pm304.5$ & $4.7\pm6.8$ & \cellcolor{green!15} $0.0190\pm0.0109$ & \cellcolor{green!15} $0.0099\pm0.0065$ & $1.034\pm0.179$ & $0.602\pm0.265$ & $1.99\pm1.66$ & $75.7\pm15.6$ & $68.1\pm17.5$ & $69.3\pm12.2$ & $0.405$ \\
\hline
\end{tabular}%
}
\label{tab:eval_small_models}
\end{table*}

\begin{table*}[!t]
\centering
\caption{Evaluation on one-kit and all-kits test sets (Large models).}
\renewcommand{\arraystretch}{1}
\setlength{\tabcolsep}{1pt}
\resizebox{\textwidth}{!}{%
\begin{tabular}{|p{0.060\linewidth}|p{0.060\linewidth}|p{0.060\linewidth}|p{0.060\linewidth}|p{0.060\linewidth}|p{0.07\linewidth}|p{0.07\linewidth}|p{0.11\linewidth}|p{0.11\linewidth}|p{0.12\linewidth}|p{0.067\linewidth}|p{0.067\linewidth}|p{0.067\linewidth}|p{0.064\linewidth}|}
\hline
\textbf{Eval} & \textbf{Codec} & \multicolumn{3}{c|}{\textbf{Token metrics}} & \multicolumn{5}{c|}{\textbf{Audio metrics}} & \multicolumn{3}{c|}{\textbf{Onset metrics}} & \\
\cline{3-5}\cline{6-10}\cline{11-13}
\textbf{Setting} & \textbf{} & \shortstack[c]{\scriptsize \textbf{\textit{NLL}$^{\mathrm{a}}$\,\,$\downarrow$}} & \shortstack[c]{\scriptsize \textbf{\textit{PPL}$^{\mathrm{a}}$\,\,$\downarrow$}} & \shortstack[c]{\scriptsize \textbf{\textit{Acc(\%)}$^{\mathrm{a}}$\,\,$\uparrow$}} & \shortstack[c]{\scriptsize \textbf{\textit{RMSE}$^{\mathrm{b}}$\,\,$\downarrow$}} & \shortstack[c]{\scriptsize \textbf{\textit{MAE}$^{\mathrm{b}}$\,\,$\downarrow$}} & \shortstack[c]{\scriptsize \textbf{\textit{MR-STFT SC}$^{\mathrm{b}}$\,\,$\downarrow$}} & \shortstack[c]{\scriptsize \textbf{\textit{Env RMS corr}$^{\mathrm{b}}$\,\,$\uparrow$}} & \shortstack[c]{\scriptsize \textbf{\textit{TTER (dB) MAE}$^{\mathrm{b}}$\,\,$\downarrow$}} & \shortstack[c]{\scriptsize \textbf{\textit{P(\%)}$^{\mathrm{c}}$\,\,$\uparrow$}} & \shortstack[c]{\scriptsize \textbf{\textit{R(\%)}$^{\mathrm{c}}$\,\,$\uparrow$}} & \shortstack[c]{\scriptsize \textbf{\textit{F1(\%)}$^{\mathrm{c}}$\,\,$\uparrow$}} & \shortstack[c]{\scriptsize \textbf{\textit{FAD}$^{\mathrm{d}}$\,\,$\downarrow$}} \\
\hline
 OneKit & encodec & \cellcolor{green!15} $2.847\pm0.667$ & \cellcolor{green!15} $22.0\pm18.4$ & \cellcolor{green!15} $36.9\pm14.4$ & $0.0167\pm0.0086$ & $0.0082\pm0.0048$ & $0.997\pm0.004$ & $0.064\pm0.203$ & $4.66\pm2.00$ & $37.0\pm17.4$ & \cellcolor{green!15} $17.3\pm7.1$ & \cellcolor{green!15} $23.0\pm9.0$ & \cellcolor{green!15} $0.972$ \\
\hline
 OneKit & xcodec & $5.399\pm0.196$ & $225.8\pm49.2$ & $5.3\pm1.5$ & $0.0167\pm0.0086$ & $0.0083\pm0.0048$ & \cellcolor{green!15} $0.996\pm0.005$ & \cellcolor{green!15} $0.098\pm0.172$ & \cellcolor{green!15} $3.27\pm1.98$ & $12.0\pm32.5$ & $1.1\pm3.9$ & $2.0\pm6.0$ & $1.090$ \\
\hline
 OneKit & dac & $6.518\pm0.177$ & $686.4\pm99.0$ & $2.4\pm5.1$ & \cellcolor{green!15} $0.0167\pm0.0086$ & \cellcolor{green!15} $0.0082\pm0.0048$ & $0.998\pm0.002$ & $-0.130\pm0.154$ & $3.32\pm1.98$ & \cellcolor{green!15} $84.5\pm26.6$ & $13.6\pm6.5$ & $22.9\pm9.5$ & $1.250$ \\
\hline
\hline
 AllKits & encodec & \cellcolor{green!15} $2.966\pm0.811$ & \cellcolor{green!15} $29.0\pm46.8$ & \cellcolor{green!15} $38.1\pm15.1$ & $0.0161\pm0.0094$ & $0.0083\pm0.0057$ & $0.996\pm0.006$ & $0.052\pm0.218$ & $4.78\pm2.40$ & $37.0\pm17.4$ & \cellcolor{green!15} $17.3\pm7.1$ & $23.0\pm9.0$ & \cellcolor{green!15} $0.943$ \\
\hline
 AllKits & xcodec & $5.535\pm0.224$ & $259.8\pm59.1$ & $5.5\pm1.9$ & $0.0161\pm0.0094$ & $0.0084\pm0.0057$ & \cellcolor{green!15} $0.994\pm0.011$ & \cellcolor{green!15} $0.073\pm0.185$ & $3.59\pm2.39$ & $69.1\pm23.8$ & $16.1\pm7.4$ & \cellcolor{green!15} $25.4\pm9.5$ & $1.099$ \\
\hline
 AllKits & dac & $6.502\pm0.221$ & $680.5\pm122.1$ & $2.9\pm5.8$ & \cellcolor{green!15} $0.0161\pm0.0094$ & \cellcolor{green!15} $0.0083\pm0.0057$ & $0.998\pm0.002$ & $-0.110\pm0.159$ & \cellcolor{green!15} $3.45\pm2.38$ & \cellcolor{green!15} $84.5\pm26.6$ & $13.6\pm6.5$ & $22.9\pm9.5$ & $1.200$ \\
\hline
\end{tabular}%
}
\vspace{2pt}
\parbox{\linewidth}{\footnotesize $^{\mathrm{a}}$PAD ignored; mean$\pm$std over windows. $^{\mathrm{b}}$Tokens decoded then resampled to 32\,kHz; mean$\pm$std over windows. $^{\mathrm{c}}$Onset metrics match predicted-audio onsets to grid-derived GT onsets within 50\,ms (GT velocity$>0.30$). $^{\mathrm{d}}$fadtk embedding: CLAP-LAION-music; per-run variant: OneKit: FAD$\infty$, clap-laion-music, n=1748; AllKits: FAD, clap-laion-music, n=68180}
\label{tab:eval_big_models}
\end{table*}

\section{Results}
Tables~\ref{tab:eval_small_models} and~\ref{tab:eval_big_models} report token prediction performance and downstream audio quality for EnCodec, DAC, and X-Codec under single-kit (OneKit) and multi-kit (AllKits) evaluation. Across both settings, EnCodec is the easiest token space to model and achieves the best overall trade-off between token-level accuracy and audio-level quality. In contrast, scaling the Transformer from \textit{Base} to \textit{Large} consistently degrades performance across codecs, indicating optimization instability in our current training recipe.

\subsection{Base models}
For the \textit{Base} models (Table~\ref{tab:eval_small_models}), EnCodec achieves the lowest token NLL/PPL and the highest token accuracy in both settings (OneKit: $42.7\%$; AllKits: $43.4\%$). This advantage largely carries over to audio-space metrics: EnCodec attains the best MR-STFT spectral convergence and the highest RMS-envelope correlation in both settings, and it yields the lowest perceptual distance (FAD: $0.281$ on OneKit; $0.193$ on AllKits). DAC attains the lowest sample-level error (RMSE/MAE), but token prediction is markedly harder (very high PPL and low accuracy), and DAC has the worst FAD in both settings (OneKit: $0.545$; AllKits: $0.405$). X-Codec sits between these extremes, with substantially worse token metrics than EnCodec and correspondingly worse audio metrics and FAD (e.g., OneKit FAD $0.350$; AllKits FAD $0.277$).

\subsection{Large models}
Increasing capacity (\textit{Large} models, Table~\ref{tab:eval_big_models}) does not improve results; instead, all codecs degrade. Token metrics worsen (higher NLL/PPL, lower accuracy) and audio metrics indicate failure modes consistent with unstable training: envelope correlation collapses toward zero or becomes negative (e.g., OneKit EnCodec $0.064$; DAC $-0.130$), transient-to-tail error increases (e.g., EnCodec TTER MAE $\approx 4.7$\,dB), and onset F1 drops sharply for EnCodec and X-Codec (typically $\approx 2$--$25\%$ depending on setting). The perceptual metric mirrors these failures, with substantially higher FAD for all codecs under the \textit{Large} configuration (e.g., OneKit: EnCodec $0.972$, X-Codec $1.090$, DAC $1.250$).

\subsection{Base models on all-kits}
To assess variability across kits, we rank kits by EnCodec token accuracy for the \textit{Base} all-kits model and report the top-3 and bottom-3 kits in Table~\ref{tab:top_worst_kits_allkits_compact}. We restrict this analysis to \textit{Base} because the \textit{Large} models are unreliable. While EnCodec remains the most learnable token space overall, per-kit results also show that higher token accuracy is not a sufficient proxy for audio quality; the full per-kit breakdown is available on the project page under the \texttt{plots} folder.


\begin{table*}[htbp]
\caption{Top-3 and worst-3 kits by EnCodec token accuracy (all-kits). Best values per kit highlighted.}
\begin{center}
\renewcommand{\arraystretch}{0.95}
\setlength{\tabcolsep}{0.6pt}
\scriptsize
\resizebox{\textwidth}{!}{%
\begin{tabular}{|p{0.055\linewidth}|p{0.165\linewidth}|p{0.055\linewidth}|p{0.050\linewidth}|p{0.050\linewidth}|p{0.055\linewidth}|p{0.060\linewidth}|p{0.060\linewidth}|p{0.070\linewidth}|p{0.070\linewidth}|p{0.070\linewidth}|p{0.052\linewidth}|p{0.052\linewidth}|p{0.052\linewidth}|p{0.055\linewidth}|}
\hline
\textbf{Eval} & \textbf{Kit} & \textbf{Codec} & \multicolumn{3}{c|}{\textbf{Token}} & \multicolumn{5}{c|}{\textbf{Audio}} & \multicolumn{3}{c|}{\textbf{Onset}} & \\
\cline{4-6}\cline{7-11}\cline{12-14}
\textbf{Set} & \textbf{} & \textbf{} & \shortstack[c]{\scriptsize \textbf{NLL}$^{\mathrm{a}}$\,\,$\downarrow$} & \shortstack[c]{\scriptsize \textbf{PPL}$^{\mathrm{a}}$\,\,$\downarrow$} & \shortstack[c]{\scriptsize \textbf{Acc(\%)}$^{\mathrm{a}}$\,\,$\uparrow$} & \shortstack[c]{\scriptsize \textbf{RMSE}$^{\mathrm{b}}$\,\,$\downarrow$} & \shortstack[c]{\scriptsize \textbf{MAE}$^{\mathrm{b}}$\,\,$\downarrow$} & \shortstack[c]{\scriptsize \textbf{MRSTFT}$^{\mathrm{b}}$\,\,$\downarrow$} & \shortstack[c]{\scriptsize \textbf{Env}$^{\mathrm{b}}$\,\,$\uparrow$} & \shortstack[c]{\scriptsize \textbf{TTER}$^{\mathrm{b}}$\,\,$\downarrow$} & \shortstack[c]{\scriptsize \textbf{P(\%)}$^{\mathrm{c}}$\,\,$\uparrow$} & \shortstack[c]{\scriptsize \textbf{R(\%)}$^{\mathrm{c}}$\,\,$\uparrow$} & \shortstack[c]{\scriptsize \textbf{F1(\%)}$^{\mathrm{c}}$\,\,$\uparrow$} & \shortstack[c]{\scriptsize \textbf{FAD}$^{\mathrm{d}}$\,\,$\downarrow$} \\
\hline
 \textbf{Top} & Shuffle (Blues) & dac & $6.17$ & $521.7$ & $4.8$ & \cellcolor{green!15} $0.0131$ & \cellcolor{green!15} $0.0063$ & $1.080$ & $0.68$ & $2.46$ & $78$ & \cellcolor{green!15} $67$ & \cellcolor{green!15} $70$ & $0.449$ \\
  &  & encodec & \cellcolor{green!15} $1.83$ & \cellcolor{green!15} $7.9$ & \cellcolor{green!15} $50.5$ & $0.0134$ & $0.0066$ & \cellcolor{green!15} $0.826$ & \cellcolor{green!15} $0.74$ & \cellcolor{green!15} $1.40$ & \cellcolor{green!15} $83$ & $58$ & $66$ & \cellcolor{green!15} $0.332$ \\
  &  & xcodec & $4.49$ & $109.8$ & $12.1$ & $0.0252$ & $0.0139$ & $2.072$ & $0.47$ & $2.10$ & $76$ & $66$ & $68$ & $0.407$ \\
\hline
 \textbf{Top} & Warmer Funk & dac & $5.89$ & $429.1$ & $7.6$ & \cellcolor{green!15} $0.0150$ & \cellcolor{green!15} $0.0061$ & $0.996$ & $0.65$ & $1.86$ & $75$ & $70$ & $71$ & $0.408$ \\
  &  & encodec & \cellcolor{green!15} $1.89$ & \cellcolor{green!15} $9.0$ & \cellcolor{green!15} $50.2$ & $0.0164$ & $0.0068$ & \cellcolor{green!15} $0.847$ & \cellcolor{green!15} $0.70$ & \cellcolor{green!15} $1.85$ & $77$ & \cellcolor{green!15} $73$ & \cellcolor{green!15} $73$ & \cellcolor{green!15} $0.253$ \\
  &  & xcodec & $4.44$ & $110.4$ & $12.2$ & $0.0267$ & $0.0107$ & $1.566$ & $0.64$ & $1.96$ & \cellcolor{green!15} $79$ & $62$ & $68$ & $0.310$ \\
\hline
 \textbf{Top} & 60s Rock & dac & $5.95$ & $444.1$ & $7.4$ & \cellcolor{green!15} $0.0139$ & \cellcolor{green!15} $0.0065$ & $0.973$ & $0.65$ & $1.83$ & $77$ & $67$ & $69$ & $0.346$ \\
  &  & encodec & \cellcolor{green!15} $1.84$ & \cellcolor{green!15} $8.1$ & \cellcolor{green!15} $49.9$ & $0.0153$ & $0.0071$ & \cellcolor{green!15} $0.844$ & \cellcolor{green!15} $0.72$ & \cellcolor{green!15} $1.61$ & $80$ & \cellcolor{green!15} $68$ & \cellcolor{green!15} $72$ & \cellcolor{green!15} $0.266$ \\
  &  & xcodec & $4.38$ & $103.0$ & $13.1$ & $0.0222$ & $0.0100$ & $1.274$ & $0.65$ & $1.83$ & \cellcolor{green!15} $81$ & $66$ & $71$ & $0.329$ \\
\hline
\hline\hline
 \textbf{Worst} & Classic Rock & dac & $6.33$ & $610.1$ & $3.3$ & $0.0234$ & \cellcolor{green!15} $0.0132$ & $1.023$ & $0.62$ & $2.64$ & $72$ & \cellcolor{green!15} $66$ & $66$ & $0.499$ \\
  &  & encodec & \cellcolor{green!15} $2.63$ & \cellcolor{green!15} $19.3$ & \cellcolor{green!15} $34.5$ & \cellcolor{green!15} $0.0226$ & $0.0132$ & \cellcolor{green!15} $0.788$ & \cellcolor{green!15} $0.72$ & \cellcolor{green!15} $1.18$ & $77$ & $65$ & \cellcolor{green!15} $68$ & \cellcolor{green!15} $0.197$ \\
  &  & xcodec & $4.52$ & $112.8$ & $12.0$ & $0.0450$ & $0.0266$ & $2.177$ & $0.40$ & $2.82$ & \cellcolor{green!15} $77$ & $61$ & $66$ & $0.455$ \\
\hline
 \textbf{Worst} & Arena Stage & dac & $6.32$ & $593.7$ & $3.1$ & $0.0265$ & $0.0159$ & $1.153$ & $0.50$ & $2.65$ & $75$ & $68$ & $69$ & $0.489$ \\
  &  & encodec & \cellcolor{green!15} $2.59$ & \cellcolor{green!15} $18.3$ & \cellcolor{green!15} $34.5$ & \cellcolor{green!15} $0.0241$ & \cellcolor{green!15} $0.0146$ & \cellcolor{green!15} $0.780$ & \cellcolor{green!15} $0.70$ & \cellcolor{green!15} $1.63$ & $78$ & \cellcolor{green!15} $69$ & \cellcolor{green!15} $70$ & \cellcolor{green!15} $0.155$ \\
  &  & xcodec & $4.44$ & $107.0$ & $12.2$ & $0.0404$ & $0.0247$ & $1.604$ & $0.49$ & $2.67$ & \cellcolor{green!15} $79$ & $66$ & $69$ & $0.334$ \\
\hline
 \textbf{Worst} & Ele-Drum & dac & $6.05$ & $483.3$ & $5.9$ & \cellcolor{green!15} $0.0250$ & \cellcolor{green!15} $0.0143$ & $0.986$ & $0.70$ & $1.62$ & $70$ & $67$ & $66$ & $0.325$ \\
  &  & encodec & \cellcolor{green!15} $2.46$ & \cellcolor{green!15} $14.7$ & \cellcolor{green!15} $35.9$ & $0.0262$ & $0.0149$ & \cellcolor{green!15} $0.834$ & \cellcolor{green!15} $0.75$ & \cellcolor{green!15} $1.59$ & $74$ & $66$ & $67$ & $0.187$ \\
  &  & xcodec & $4.27$ & $89.8$ & $12.9$ & $0.0581$ & $0.0355$ & $2.486$ & $0.54$ & $1.88$ & \cellcolor{green!15} $76$ & \cellcolor{green!15} $68$ & \cellcolor{green!15} $70$ & \cellcolor{green!15} $0.155$ \\
\hline
\end{tabular}%
}
\vspace{1pt}
\parbox{\linewidth}{\scriptsize $^{\mathrm{a}}$PAD ignored. $^{\mathrm{b}}$Decoded audio at 32\,kHz. $^{\mathrm{c}}$Onsets: pred-audio vs grid GT within 50\,ms (GT vel$>0.30$). $^{\mathrm{d}}$Per-kit FAD from \texttt{fadtk} (CLAP-LAION-music).}
\label{tab:top_worst_kits_allkits_compact}
\end{center}
\end{table*}

\section{Discussion}
We presented \emph{DrumGrid2Audio}, a conditional drum renderer that predicts neural codec tokens from expressive drum grids and decodes them to waveform with a fixed pretrained codec. Under a common non-autoregressive Transformer predictor trained on E-GMD, EnCodec was consistently the most learnable target representation and achieved the strongest overall audio quality across both OneKit and AllKits, whereas DAC and X-Codec were substantially harder to model and tended to underperform on perceptual metrics even when sample-level errors were competitive. We also observed that na\"ively increasing model capacity can be counterproductive in this setting: the \textit{Large} models frequently exhibited unstable optimization and degraded audio quality, motivating future work on more robust training and scaling behavior.

As a coarse diagnostic of the token spaces, we observed that EnCodec tends to exhibit a more constrained and redundant token stream on our drum windows---e.g., lower effective entropy and slower token turnover than DAC or X-Codec. While these statistics are not sufficient to predict audio quality on their own, they suggest that the conditional mapping $p(\text{tokens}\mid\text{grid})$ may be comparatively easier to learn for EnCodec, providing a plausible partial explanation for its consistently stronger performance in our experiments. The corresponding diagnostic results are included in the project’s \texttt{plots/} directory. 

Looking forward, two practical directions are (i) a \emph{humanization pipeline}, where a symbolic front-end generates coarse or quantized drum grids (e.g., style- and density-controlled patterns) and DrumGrid2Audio renders them into human-like audio with microtiming, dynamics, and articulation-dependent timbre, and (ii) a \emph{knob-conditioned renderer} that takes a simpler grid representation together with explicit control variables for expressivity (e.g., per-instrument hit-rate, velocity statistics, or timing-swing parameters) and generates audio consistent with both the rhythm and the requested expressive profile. Such controllability can be evaluated by checking whether synthesized audio matches the target statistics (via onset detection and recovered dynamics) while maintaining perceptual quality (e.g., FAD) and onset alignment.

Finally, our evaluation relies on objective and embedding-based metrics without a listening study or formal statistical testing; accordingly, the reported differences should be interpreted as descriptive observations rather than statistically significant effects. Controlled subjective evaluations remain important future work to assess perceptual realism and groove quality.


\end{document}